\documentclass[aip,jcp,reprint]{revtex4-1}
\usepackage{amsmath}
\usepackage{braket}
\usepackage{tikz}
\usepackage{multirow}
\begin{document}
\title{Longitudinal static optical properties of hydrogen chains: finite field extrapolations of matrix product state calculations}

\author{Sebastian Wouters}
\affiliation{Center for Molecular Modeling, Ghent University, Ghent, Belgium}

\author{Peter A. Limacher}
\affiliation{Department of Chemistry, McMaster University, Hamilton, Ontario, Canada}

\author{Dimitri Van Neck}
\affiliation{Center for Molecular Modeling, Ghent University, Ghent, Belgium}

\author{Paul W. Ayers}
\affiliation{Department of Chemistry, McMaster University, Hamilton, Ontario, Canada}

\begin{abstract}
We have implemented the sweep algorithm for the variational optimization of SU(2) $\otimes$ U(1) (spin and particle number) invariant matrix product states (MPS) for general spin and particle number invariant fermionic Hamiltonians. This class includes non-relativistic quantum chemical systems within the Born-Oppenheimer approximation. High-accuracy ab-initio finite field results of the longitudinal static polarizabilities and second hyperpolarizabilities of one-dimensional hydrogen chains are presented. This allows to assess the performance of other quantum chemical methods. For small basis sets, MPS calculations in the saturation regime of the optical response properties can be performed. These results are extrapolated to the thermodynamic limit.
\end{abstract}


\maketitle

\section{introduction}
Non-linear optical (NLO) properties of materials are of interest to experiment, theory, and industry. They account for a wide variety of phenomena such as frequency doubling, optical control of the refractive index, and phase conjugation.\cite{nlobook} Especially the NLO properties of linearly conjugated organic polymer-chains have moved to the center of attention and many theoretical studies have been published about the interplay of molecular structure, electron delocalization, and NLO properties.\cite{marder93,da94,doi:10.1021/cr00025a008,tykwinski98,champagne99,kirtman00,perpete08,borini09,champagne11} An important question in many of these studies is the suitability and accuracy of different quantum chemical (QC) methods,\cite{sekino08,l09} henceforth called levels of theory (LOT). Conventional density functional theory was found to dramatically overestimate NLO properties of long molecular chains.\cite{kirtman98,kirtman99} Newly developed approaches were presented to mitigate, but not fully resolve the problem.\cite{oep03,sekino05} In the meantime also certain irregularities between Hartree-Fock (HF) and post-HF methods were noticed, calling into question the importance and the influence of electron correlation on NLO properties.\cite{toto95,li08,2011JChPh.135a4111L} It is therefore desirable to obtain the NLO properties of the fully correlated problem, i.e. at exact diagonalization (ED) accuracy. Linear chains of hydrogen are ideal test systems for assessing the quality of different LOTs.\cite{PhysRevA.52.178,PhysRevA.52.1039,sekino07,2009IJQC..109.3103C}

A recently developed class of variational ansatzes, the tensor network states (TNS), yield compact and accurate approximations of low-lying eigenstates based on the topological properties of the Hamiltonian. The matrix product state (MPS) is the natural TNS for one-dimensional holographic geometries.\cite{2011JSP142E} Conversely, it can be shown that every quantum many-body state can be rewritten as an MPS.\cite{schollwockDMRGatMPSage} This allows the MPS to be used as a variational ansatz for any quantum system. The optimal MPS can be found implicitly by means of the density matrix renormalization group (DMRG) or explicitly by variationally optimizing the MPS.\cite{PhysRevB.55.2164,schollwockDMRGatMPSage} Several groups have implemented the DMRG algorithm for ab-initio QC calculations.\cite{1999JChPh.110.4127W, Daulpaper, 2001JChPh.115.6815M, 2002JChPh.116.4462C, PhysRevB.67.125114, PhysRevB.68.195116, PhysRevB.70.205118, 2005JChPh.122b4107M, 2006CP323519R, JCPmortiz126, 2008JChPh.128a4107Z, 2009JChPh.130w4114K, PhysRevB.81.235129} For quasi-one-dimensional chemical systems like hydrogen chains,\cite{2006JChPh.125n4101H} the MPS gives an efficient description. The mutual screening of electrons and nuclei results in an effectively local electromagnetic interaction, which explains why DMRG works well for these systems.\cite{Daulpaper, 2009JChPh.130w4114K} For systems that don't have a one-dimensional holographic geometry, the MPS is not always efficient, as can be seen by the virtual dimensions required to obtain near-ED accuracy.\cite{1999JChPh.110.4127W,2009JChPh.130w4114K} A better choice and ordering of the single particle basis can resolve the problem partly.\cite{PhysRevB.82.205105, PhysRevA.83.012508, 1999JChPh.110.4127W, 2006JChPh.124c4103M, JCPmortiz126, 2002JChPh.116.4462C, PhysRevB.68.195116, PhysRevB.70.205118} Other TNSs such as the tree TNS\cite{PhysRevB.82.205105,PhysRevA.83.012508} or different ansatzes such as correlator product states\cite{2009NJPh...11h3026M, PhysRevB.80.245116} (e.g. the complete-graph TNS\cite{2010NJPh12j3008M}) can further improve the descriptions of such systems. It has even been suggested to use correlator product states with auxiliary indices.\cite{PhysRevB.80.245116} This leads us back to White's original proposal\cite{1999JChPh.110.4127W} to combine several orbitals into a single local degree of freedom in QC DMRG.

Together with an efficient TNS, the use of symmetry can make the description of eigenstates even more compact. Structuring the virtual bonds according to the irreducible representations of the applied symmetry groups introduces a sparse block structure in the tensors. For non-Abelian symmetry groups the Wigner-Eckart theorem permits working with reduced tensors.\cite{PhysRevA.82.050301, 2010NJPh12c3029S, 2007JSMTE1014M, 2002EL57852M}

In this paper, we use the SU(2) $\otimes$ U(1) invariant MPS to study the longitudinal static dipole polarizability and second hyperpolarizability of one-dimensional hydrogen chains by means of finite field extrapolations. The MPS algorithm enables us to study longer chains than with ED, but not at the expense of decreasing accuracy. For small basis sets, this allows us to obtain high-accuracy data in the saturation regime of the optical response properties. The results obtained with our MPS algorithm let us assess the performance of standard QC methods. When possible, these results are extrapolated to infinite chain length to obtain quantative results in the thermodynamic (TD) limit. Different basis sets are compared.

Related work, studying both the static and dynamic polarizabilities and second hyperpolarizabilities of conjugated $\pi$-systems, includes the analytic response theory for ab-initio QC DMRG\cite{2009JChPh.130r4111D} and the correction vector DMRG algorithm for Pariser-Parr-Pople Hamiltonians.\cite{mukhopadhyay2009study, 2011arXiv1108.0854K} Accurate TD limit data of the static optical response properties of hydrogen chains can also be obtained with diffusion Monte Carlo, using the modern theory of polarization.\cite{PhysRevLett.95.207602, 2009JChPh.131i4104U}

The MPS ansatz is briefly addressed in section \ref{MPSlagosection}, where the variational optimization of the MPS for ab-initio QC Hamiltonians, imposing SU(2) $\otimes$ U(1) spin and particle number symmetry, and our implementation are also discussed. The finite field method is outlined in section \ref{theFFmethod}. Section \ref{equallyspacedHchain} deals with the optical properties of several spin states of an equally spaced hydrogen chain, where the spacing controls the amount of static correlation. A chain of $\text{H}_2$ constituents is studied in section \ref{H2dimerssection}: the influence of intermolecular distance (and hence the amount of electron delocalization), LOT and basis set on the optical properties are determined. When possible, the MPS results are extrapolated to the TD limit. Section \ref{sectconclusions} contains the conclusions.

\section{The MPS algorithm \label{MPSlagosection}}
As there are already excellent works on the variational optimization of an MPS,\cite{schollwockDMRGatMPSage} on the implementation of DMRG for ab-initio QC calculations,\cite{1999JChPh.110.4127W, Daulpaper, 2001JChPh.115.6815M, 2002JChPh.116.4462C, PhysRevB.67.125114, PhysRevB.68.195116, PhysRevB.70.205118, 2005JChPh.122b4107M, 2006CP323519R, JCPmortiz126, 2008JChPh.128a4107Z, 2009JChPh.130w4114K, PhysRevB.81.235129, 2004JChPh.120.3172C} and on the use of non-Abelian symmetries in TNSs,\cite{PhysRevA.82.050301, 2010NJPh12c3029S, 2002EL57852M, 2007JSMTE1014M} we choose to focus only on how these principal concepts contribute to our algorithm.

\subsection{DMRG and MPS}
In non-relativistic ab-initio QC, the positions of the nuclei are fixed in the Born-Oppenheimer approximation and a basis set is chosen as the orbital degrees of freedom. Because we study one-dimensional systems in this work, L\"owdin transformed Gaussian basis sets are used as they preserve locality well.\cite{2006JChPh.125n4101H, 1957PhRv105102C} Consider a state with $L$ orbitals and 4 possible occupations $i$ per orbital:
\begin{equation}
\ket{\Psi} = \sum\limits_{\{i_1 ... i_L\}} c_{i_1 ... i_L} \ket{i_1 ... i_L}
\end{equation}
This state can always be rewritten as an MPS:\cite{schollwockDMRGatMPSage}
\begin{equation}
\ket{\Psi} = \sum\limits_{\{i_1 ... i_L\}} \sum\limits_{\{k_1 ... k_{L-1}\}} M^{i_1}_{k_1} M^{i_2}_{k_1 k_2} ... M^{i_L}_{k_{L-1}} \ket{i_1 ... i_L} \label{MPSnotationeq}
\end{equation}
which associates to every orbital 4 matrices $M^i_{k_Lk_R}$ or a single three-index tensor. The index $i$ is called the local index and represents the occupation. The indices $k_L$ and $k_R$ are called virtual indices. The dimension $D$ of the virtual indices needs to increase exponentially towards the middle of the MPS chain for \eqref{MPSnotationeq} to represent the full Hilbert space. In calculations, the virtual dimension $D$ is truncated and the MPS represents only a part of the full Hilbert space. The tensors in the MPS chain are iteratively optimized, one at a time, in the sweep algorithm.\cite{schollwockDMRGatMPSage} This method is strictly variational. For arbitrarily large systems with a one-dimensional holographic geometry, the ED solution can be approximated to any desired accuracy by an MPS with a finite $D$.\cite{2011JSP142E} Note that Eq. \eqref{MPSnotationeq} represents a multideterminantal wavefunction and is hence able to capture static correlation.\cite{2006JChPh.125n4101H}

There are two versions of the DMRG algorithm: single-site and two-site DMRG. Their names refer to the number of neighbouring orbitals that are free at a local optimization step. The variational optimization of an MPS corresponds to (but is not equal to) single-site DMRG. In the MPS algorithm the renormalization transformations and subsequent decimations of the DMRG algorithm are incorporated in the MPS ansatz itself. Fixed points of both DMRG algorithms can be written as MPSs.\cite{PhysRevB.55.2164} Single-site DMRG is also strictly variational, while two-site DMRG is not.\cite{2002JChPh.116.4462C, 2008JChPh.128n4115Z}

In certain cases the two-site DMRG algorithm and the variational optimization of the corresponding MPS both lead to the same result. This is often the case for systems that have one-dimensional holographic geometries and for which the MPS is the natural TNS, while for other systems the two-site DMRG algorithm can outperform the single-site variational optimization of an MPS as it provides more degrees of freedom for each local diagonalization step.\cite{schollwockDMRGatMPSage, PhysRevB.67.125114} In both DMRG algorithms, adding perturbative corrections or noise to the reduced density matrix helps to reach the true ground state within the subspace of the full Hilbert space spanned by the MPS, as they help to reintroduce lost quantum numbers in the reduced basis.\cite{schollwockDMRGatMPSage, 2002JChPh.116.4462C, 1999JChPh.110.4127W, 2009JChPh.130w4114K, PhysRevB.72.180403, 2008JChPh.128n4115Z} Another way to achieve this, is to explicitly keep states with a certain symmetry in the reduced basis.\cite{2004JChPh.120.3172C}

\begin{figure}
\includegraphics[width=0.40\textwidth]{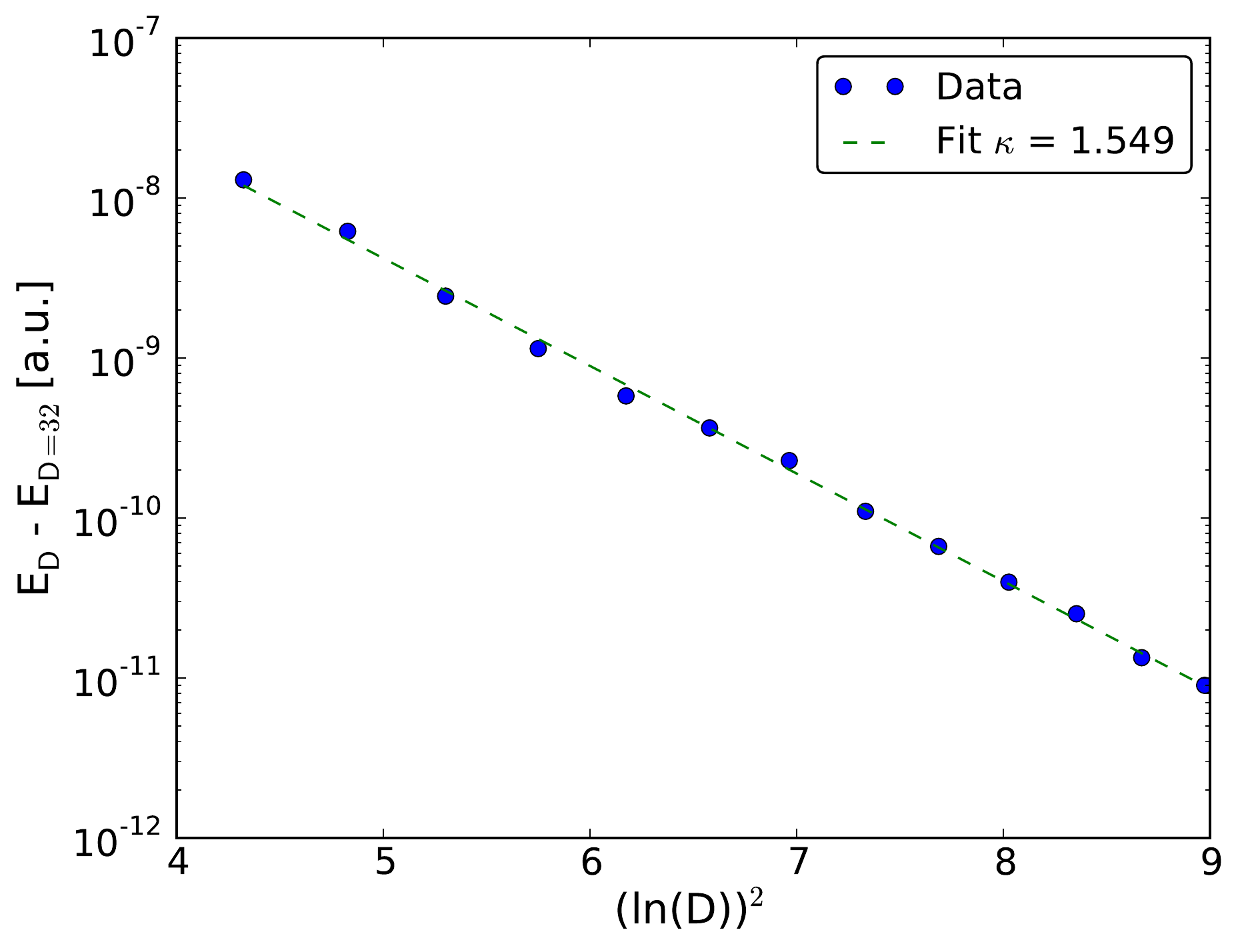}
\caption{\label{ConvergencePlot} The ground state of a hydrogen chain of 36 atoms, with an alternate atom spacing of 2/3 a.u. (see section \ref{secConfigDimerExplanation}), in the L\"owdin transformed STO-6G basis, is approximated by several MPSs with increasing virtual dimension. The scaling of the ground state energy with $D$, the virtual dimension \textit{per symmetry sector} (see section \ref{secImplementation}), follows \eqref{Chanscaling}. The rightmost data point corresponds to $D=20$ and $E_{D=32}$ is used as an approximation to the exact result.}
\end{figure}

For the systems in our study, the holographic geometry is one-dimensional and hence the MPS ansatz is a good choice. This is confirmed by the rapid convergence of the ground state energy obtained with an MPS with increasing virtual dimension. Chan \textit{et al.}\cite{Chan_Ayers_Croot_2002,2002JChPh.116.4462C} have proposed a relation for this convergence:
\begin{equation}
\ln(E_{D} - E_{\text{exact}} ) = a - \kappa (\ln(D))^2 \label{Chanscaling}
\end{equation}
Here, $a$ and $\kappa$ are fitting parameters, $E_{\text{exact}}$ is the ED result and $E_D$ the energy when an MPS with virtual dimension $D$ is used. Eq. \eqref{Chanscaling} is illustrated in Fig. \ref{ConvergencePlot}.

\subsection{General two-body Hamiltonians}
In second quantization the Hamiltonian can be written as:\cite{thebible}
\begin{multline}
\hat{H}_0 = \sum\limits_{i,j,\sigma} (i | \hat{T} | j) \hat{a}^{\dagger}_{i \sigma} \hat{a}_{j \sigma} \\
+ \frac{1}{2} \sum\limits_{i,j,k,l, \sigma, \tau} (i j | \hat{V} | k l) \hat{a}^{\dagger}_{i \sigma} \hat{a}^{\dagger}_{j \tau} \hat{a}_{l \tau} \hat{a}_{k \sigma} \label{hamiltonian1}
\end{multline}
where the Latin letters denote orbitals and the Greek letters spin projections. Global spin and global particle number are conserved by this Hamiltonian. The matrix elements are calculated based on the work of Obara and Saika.\cite{1986JChPh..84.3963O}

For the local optimization procedure, partial Hamiltonian terms such as e.g. $\hat{a}^{\dagger}_{i \in \text{left} ~ \sigma} \hat{a}^{\dagger}_{j \in \text{left} ~ \tau}$ need to be stored in memory. We have taken all previous considerations in the literature into account to store as few of them as possible.\cite{2002JChPh.116.4462C, 2004JChPh.120.3172C, 2009JChPh.130w4114K} These include multiplying creators/annihilators with two-body matrix elements and contracting common indices to form complementary operators, exploiting the hermitian symmetry of matrix elements as well as exploiting the creator/annihilator swap symmetry due to the fermion anti-commutation relations. Further storage reduction is possible by exploiting global symmetry.

\subsection{Global symmetries}
Using the global symmetries of the Hamiltonian has many advantages, including the ability to explicitly scan only the desired symmetry sector of the total Hilbert space, and an improvement of computational performance. This improvement consists of a reduction in both CPU time (by reducing the number of sweeps) and memory usage (the tensors adopt a sparse block structure and the required virtual dimensions are smaller; this further decreases the CPU time).\cite{2002EL57852M} The main disadvantage is the increasing complexity of the algorithm: i.e. analytic work done beforehand and overhead in the resulting program. However, this needs to be done only once, and in many cases it doesn't outweigh the benefits.

We have implemented global spin and particle number symmetry. The U(1) particle number symmetry is an Abelian symmetry and is therefore represented by an additive quantum number.\cite{2007JSMTE1014M} Its implementation in ab-initio QC DMRG calculations is well known.\cite{2009JChPh.130w4114K, PhysRevB.68.195116} The SU(2) spin symmetry is a non-Abelian symmetry and requires recoupling.\cite{2007JSMTE1014M}

Global symmetry can be imposed by requiring that the three-index tensors $M^i_{k_Lk_R}$ in the MPS chain are irreducible tensor operators of the imposed symmetry group.\cite{PhysRevA.82.050301, 2010NJPh12c3029S, 2007JSMTE1014M, 2002EL57852M} The local and virtual bases are represented in states with the correct symmetry, i.e. spin $s$ or $j$, spin projection $s^z$ or $j^z$ and particle number $N$. The local states $i = \ket{-}$, $\ket{\uparrow}$, $\ket{\downarrow}$ or $\ket{\uparrow\downarrow}$ then correspond to resp. $i = \ket{s = 0; s^z = 0; N = 0}$, $\ket{\frac{1}{2} \frac{1}{2} 1}$, $\ket{\frac{1}{2} -\frac{1}{2} 1}$ and $\ket{0 0 2}$. Due to the Wigner-Eckart theorem, each irreducible tensor operator decomposes into a structural part and a degeneracy part $T$:
\begin{eqnarray}
M^i_{k_Lk_R} & = & M^{(s s^z N)}_{(j_L j_L^z N_L \alpha_L)(j_R j_R^z N_R \alpha_R)}\\
& = & \braket{j_L j_L^z s s^z | j_R j_R^z} \delta_{N_L + N, N_R} T^{(s N)}_{(j_L N_L \alpha_L)(j_R N_R \alpha_R)} \quad \label{tensordecomp}
\end{eqnarray}
The SU(2) symmetry is imposed by the Clebsch-Gordan coefficient and the U(1) symmetry by the particle conserving Kronecker delta. The indices $\alpha_L$ and $\alpha_R$ are used to keep track of the number of times an irreducible representation occurs at a virtual bond. If the virtual dimension of a symmetry sector is $D(j_L N_L) = \text{size}(\alpha_L)$, this would correspond to a dimension of $(2 j_L + 1) D(j_L N_L)$ in a non-symmetry adapted MPS.\cite{2002EL57852M} Global symmetry can be imposed by requiring that the left virtual index of the leftmost tensor in the MPS chain consists of 1 irreducible representation corresponding to $(j_L, N_L) = (0,0)$, while the right virtual index of the rightmost tensor consists of 1 irreducible representation corresponding to $(j_R N_R) = (S N)$, the desired global spin and particle number.

The operators
\begin{eqnarray}
\hat{b}^{\dagger}_{m} & = & \hat{a}^{\dagger}_{m} \label{creaannih1}\\
\hat{b}_{m} & = & (-1)^{\frac{1}{2}-m}\hat{a}_{-m} \label{creaannih2}
\end{eqnarray}
transform as irreducible tensor operators with spin $\frac{1}{2}$ under SU(2), with $m$ the spin projection.\cite{thebible} Because these operators are part of a doublet, it is possible to exploit the Wigner-Eckart theorem also for operators and complementary operators, and to develop a code without any spin projections or Clebsch-Gordan coefficients. Contracting terms of the type \eqref{tensordecomp} and \eqref{creaannih1}-\eqref{creaannih2} can be done by implicitly summing over the common multiplets and recoupling the local, virtual and operator spins. Examples are given in appendix A. Operators and complementary operators then formally consist of terms containing Clebsch-Gordan coefficients, particle conserving Kronecker deltas and reduced tensors. In our code, however, only the reduced tensors need to be calculated and stored. To the best of our knowledge, the global SU(2) symmetry has been implemented only once in ab-initio QC DMRG calculations.\cite{2008JChPh.128a4107Z} In this algorithm\cite{2008JChPh.128a4107Z} no use is made of the Wigner-Eckart theorem to work with reduced tensors, as is often proposed.\cite{PhysRevA.82.050301, 2010NJPh12c3029S, 2007JSMTE1014M, 2002EL57852M}

\subsection{Implementation \label{secImplementation}}
We have implemented the sweep algorithm \cite{schollwockDMRGatMPSage} to variationally optimize an SU(2) $\otimes$ U(1) invariant MPS in \texttt{C++}. Matrix operations are handled by LAPACK and BLAS. Wigner 6j symbols are calculated by the GNU Scientific Library. For the local optimization of the degeneracy part of an MPS tensor, we have chosen the Lanczos method, implemented in ARPACK. Where possible, the code is parallellized on a single node with OpenMP. No multinode parallellization (MPI) was needed for the results in this paper.

The virtual dimension is truncated per symmetry sector: if the virtual dimension $D(j_L N_L)$ of a symmetry sector $(j_L N_L)$ required to represent the full Hilbert space exceeds a predefined threshold $D$, it is set to $D$. For the results presented in this paper, $D$ is chosen large enough so that no relative energy error is larger than $10^{-11}$:
\begin{equation}
\frac{E_{D}-E_{\text{exact}}}{E_{\text{exact}}} < 10^{-11} \label{10minus11convergence}
\end{equation}
Specific choices for $D$ are mentioned when the applications are introduced. All tensors are stored in the minimum amount of memory required. Convergence is reached when both the energy and the wave function meet the following criteria:
\begin{eqnarray}
\mid E_n - E_{n-1} \mid & < & \epsilon_1\\
1 - \mid \braket{\text{MPS}_n \mid \text{MPS}_{n-1}} \mid & < & \epsilon_2
\end{eqnarray}
where $n$ is the sweep number and $\epsilon_1 = \epsilon_2 = 10^{-13}$ for the calculations presented in this paper. At the start of the algorithm, the MPS is filled with noise, but during the sweeps no noise or perturbative corrections were added. For more complex chemical systems, the orbital choice, the orbital ordering, and the initial guess play an important role for the convergence and even for the qualitative properties of the solution.\cite{PhysRevB.82.205105, PhysRevA.83.012508, 1999JChPh.110.4127W, 2006JChPh.124c4103M, JCPmortiz126, 2002JChPh.116.4462C, PhysRevB.68.195116} The holographic geometry of such systems is often far from one-dimensional. In DMRG calculations, basis states with a certain symmetry are sometimes explicitly kept in the reduced basis to avoid losing quantum numbers.\cite{2004JChPh.120.3172C} Note that the division of the virtual bonds in symmetry sectors $(j_L N_L)$ boils down to the same thing.

If there are $N$ electrons in the system, with $N\leq L$, the number of SU(2) $\otimes$ U(1) symmetry sectors in the middle of the chain is $\mathcal{O}(N^2)$. In that case we obtain for our algorithm a scaling per sweep of $\mathcal{O}(D^3 L^3 N^2 + D^2 L^4 N^2)$ in time and $\mathcal{O}(D^2 L^2 N^2)$ in memory.\cite{2002JChPh.116.4462C} For $N\geq L$, $N$ should be replaced by $(2L-N)$. Note that both the number of sweeps to reach convergence and the virtual dimension $D$ to reach a certain accuracy are smaller when global symmetry is imposed.\cite{2002EL57852M} Hachmann \textit{et al.}\cite{2006JChPh.125n4101H} present a method that makes use of the numerical negligibility of certain two-body matrix elements to obtain an algorithm that scales per sweep as $\mathcal{O}(D^3 L^2)$ in time and $\mathcal{O}(D^2 L)$ in memory. When applying global SU(2) $\otimes$ U(1) symmetry, these order estimates have to be multiplied with $\mathcal{O}(N^2)$ when $N\leq L$ or $\mathcal{O}((2L-N)^2)$ when $N\geq L$. The efficiency gain when neglecting these matrix elements comes with the cost of losing the variational character of the algorithm, because the Hamiltonian is altered. However, the error is under control. In the current version of our program, this quadratically scaling algorithm is not yet used, but we plan to implement it in the future.

\section{The finite field method \label{theFFmethod}}
When a homogeneous electric field $\vec{F}$ is applied, the electrons acquire a potential energy that depends on their position.\cite{kurtz1990calculation} The total Hamiltonian of the system becomes (atomic units):
\begin{equation}
\hat{H} = \hat{H}_0 + \vec{F}.\vec{r} \label{hami2}
\end{equation}
This total Hamiltonian still conserves global spin and global particle number.

The static polarizability $\alpha_{ij}$ and second hyperpolarizability $\gamma_{ijkl}$ tensors are resp. the first and third order derivatives of the electric dipole moment $\vec{\mu}$ with respect to the applied field $\vec{F}$, in the limit of an infinitesimal field:
\begin{eqnarray}
\alpha_{ij} & = & \left(\frac{\partial \mu_i(\vec{F})}{\partial F_j}\right)_{\vec{F} \rightarrow \vec{0}}\\
\gamma_{ijkl} & = & \left(\frac{\partial^3 \mu_i(\vec{F})}{\partial F_j \partial F_k \partial F_l}\right)_{\vec{F} \rightarrow \vec{0}}
\end{eqnarray}
All subscripts denote Cartesian components. Because the electric dipole moment $\vec{\mu}$ is minus the derivative of the total energy $E$ with respect to an applied electric field $\vec{F}$, $\alpha_{ij}$ and $\gamma_{ijkl}$ can also be obtained from:
\begin{eqnarray}
\alpha_{ij} & = & - \left(\frac{\partial^2 E(\vec{F})}{\partial F_i \partial F_j}\right)_{\vec{F} \rightarrow \vec{0}}\\
\gamma_{ijkl} & = & - \left(\frac{\partial^4 E(\vec{F})}{\partial F_i \partial F_j \partial F_k \partial F_l}\right)_{\vec{F} \rightarrow \vec{0}}
\end{eqnarray}
The energy $E(\vec{F})$ has to be evaluated with a wavefunction optimized for \eqref{hami2}. For molecules extending mainly in one spatial dimension (assume this to be the z-direction), the main contribution to these tensors comes from the longitudinal components $\alpha_{zz}$ and $\gamma_{zzzz}$. The hydrogen chains under study are in addition centrosymmetric. When the origin of the Cartesian coordinate system is chosen in the center of the chain, $E(\vec{F}) = E(-\vec{F})$ and the static longitudinal components of both quantities can be obtained with the following minimal finite difference formulae, where $\vec{F} = F \hat{z}$:
\begin{eqnarray}
\alpha_{zz}(F) & = & \left(\frac{2 E(0) - 2 E(F)}{F^2}\right)_{F \rightarrow 0} \label{FD1}\\
\gamma_{zzzz}(F) & = & \left(\frac{-6 E(0) + 8 E(F) - 2 E(2F)}{F^4}\right)_{F \rightarrow 0} \label{FD2}
\end{eqnarray}
\begin{figure*}[t]
 \includegraphics[width=0.855\textwidth]{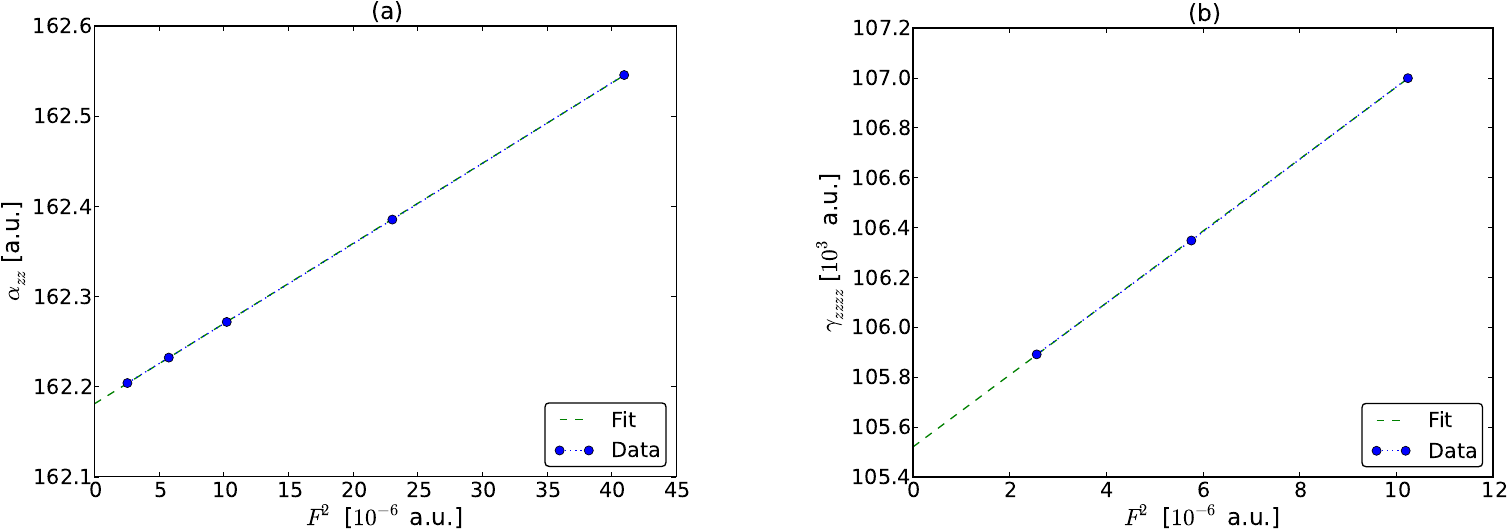}
 \caption{\label{extrapolplot} Finite field extrapolations of the static longitudinal polarizability (a) and second hyperpolarizability (b) for MPS calculations of a hydrogen chain of 36 atoms, with an alternate atom spacing of 2/3 a.u. (see section \ref{secConfigDimerExplanation}), in the L\"owdin transformed STO-6G basis. The extrapolations are done with a least squares fit to \eqref{extrapolequation}.}
\end{figure*}
The use of a finite field is explicitly incorporated in the notation: $\alpha_{zz}(F)$ and $\gamma_{zzzz}(F)$. We calculate both quantities for different values of $F$ and make a least squares extrapolation to $F=0$ according to
\begin{equation}
q(F) = q(0) + cF^2 \label{extrapolequation}
\end{equation}
where $q$ can be $\alpha_{zz}$ or $\gamma_{zzzz}$. Values of $q(0)$ and $c$ are obtained by the fit. The procedure is illustrated in Fig. \ref{extrapolplot}.

The values of $F$ are chosen with care. If they are too large, higher order effects come into play and higher order terms have to be added to \eqref{extrapolequation}. In that case, more calculations are required as more points $q(F)$ are needed to fit all parameters. Because the eigenstate energies $E_{\text{exact}}$ are approximated with MPS energies $E_D$ up to a certain accuracy, the energy differences in the numerators of \eqref{FD1} and \eqref{FD2} have a constant error. If the field values become smaller, this absolute error for the energy differences is multiplied by increasing values of $F^{-2}$ or $F^{-4}$ and the absolute error of $\alpha_{zz}(F)$ and $\gamma_{zzzz}(F)$ becomes larger. The RMS deviation of the quantities from the fit (as in Fig. \ref{extrapolplot}) will then be larger.

\begin{figure}
\includegraphics[width=0.40\textwidth]{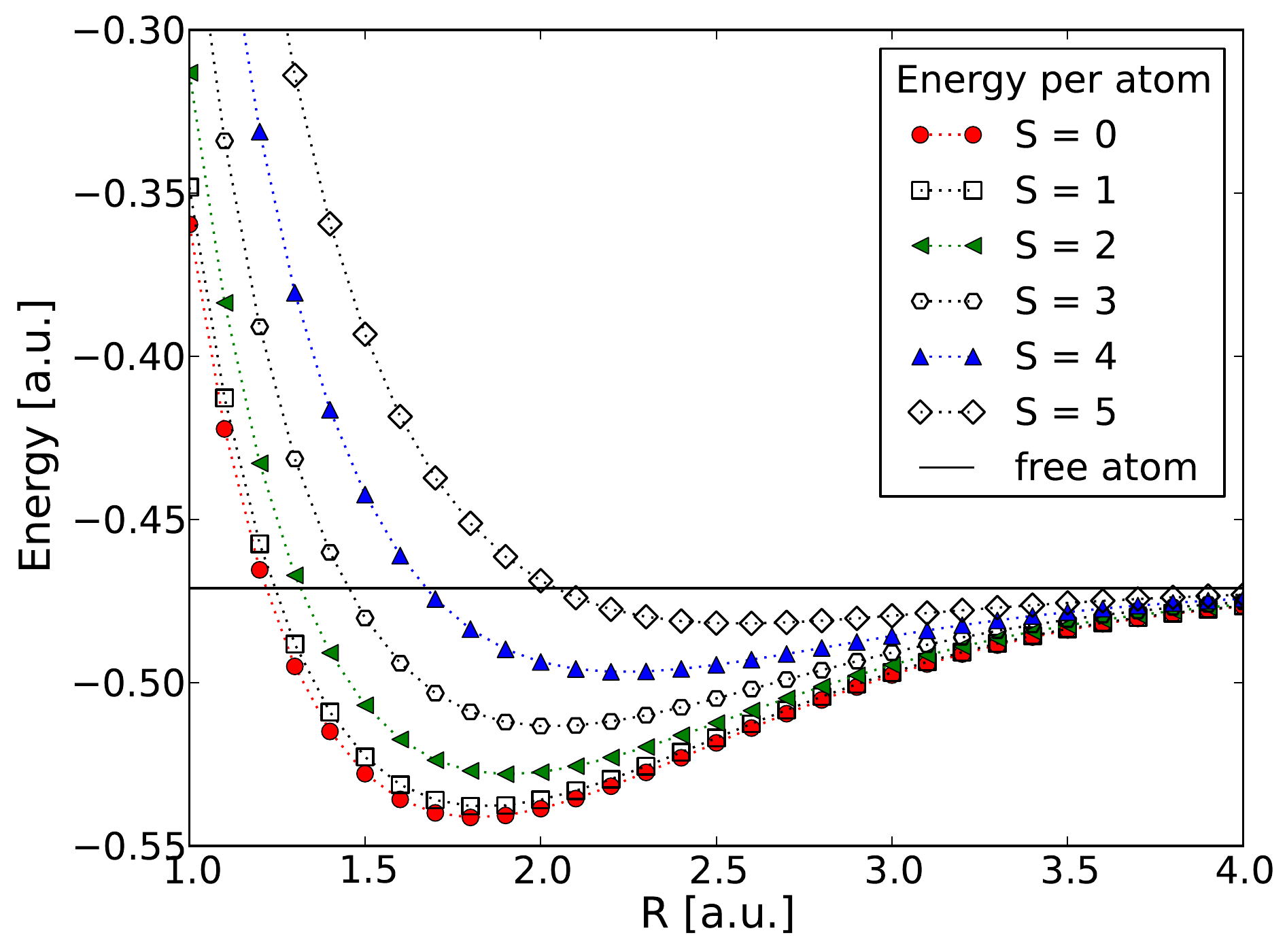}
\caption{\label{energyspinstates} The ground state energy per atom for an equally spaced hydrogen chain of 20 atoms is shown for 6 different spin states.}
\end{figure}

\begin{figure*}[t]
 \includegraphics[width=0.855\textwidth]{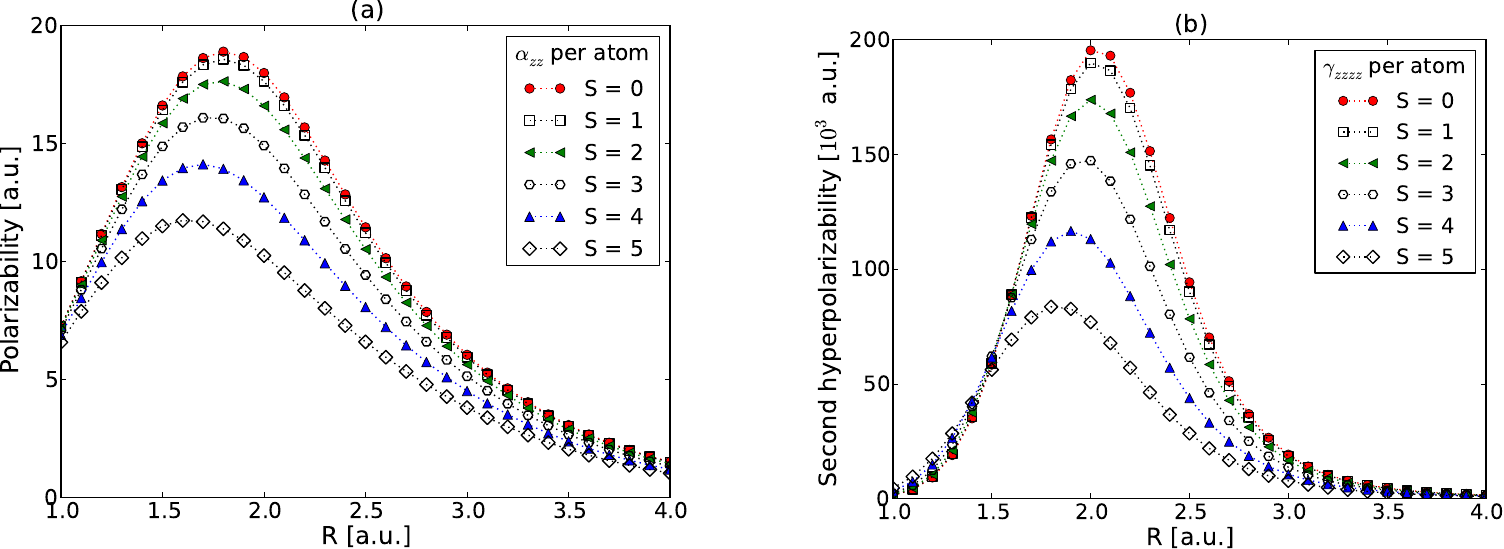}
 \caption{\label{alphaandgammaequally} The polarizabilities (a) and second hyperpolarizabilities (b) per atom for an equally spaced hydrogen chain of 20 atoms are shown for 6 different spin ground states.}
\end{figure*}

\section{Equally spaced hydrogen chain \label{equallyspacedHchain}}
Our MPS program was tested for many small systems and the results were compared with exact diagonalization (ED), confirming the correctness of our implementation. Both for this application and the next one, all presented MPS data are converged according to Eq. \eqref{10minus11convergence}.

\subsection{Introduction}
As a benchmark calculation, illustrating the possibilities of the program, the energy, as well as the static longitudinal polarizability and second hyperpolarizability of a hydrogen chain with 20 atoms are studied for different interatomic distances. The interatomic distance $R$ is defined by
\begin{equation}
\vcenter{\hbox{\setlength{\unitlength}{1cm}
\begin{picture}(6.7,1)
\put(0.1,0.375){H}
\put(0.4,0.5){\line(1,0){0.5}}
\put(1.0,0.375){H}
\put(1.3,0.5){\line(1,0){0.5}}
\put(1.9,0.375){H}
\put(2.2,0.5){\line(1,0){0.5}}
\put(2.8,0.375){H}
\put(3.1,0.5){\line(1,0){0.5}}
\put(3.7,0.375){H}
\put(4.0,0.5){\line(1,0){0.5}}
\put(4.6,0.375){H}
\put(4.9,0.5){\line(1,0){0.5}}
\put(5.5,0.375){H}
\put(5.8,0.5){\line(1,0){0.5}}
\put(6.4,0.375){H}
\put(0.5,0.6){$R$}
\put(1.4,0.6){$R$}
\put(2.3,0.6){$R$}
\put(3.2,0.6){$R$}
\put(4.1,0.6){$R$}
\put(5.0,0.6){$R$}
\put(5.9,0.6){$R$}
\end{picture}}}
\end{equation}

The study is performed for the ground states in 6 different spin symmetry sectors $S = 0,1,...,5$. The virtual dimension per symmetry sector was truncated to $D = 64$ for all the results in this section. The energies were determined for 6 field values $F = 0, 0.0008, 0.0012, 0.0016, 0.0024$ and $0.0032$ a.u., yielding 5 points for the $\alpha_{zz}$ extrapolation and 3 points for the $\gamma_{zzzz}$ extrapolation. The minimal basis set STO-6G \cite{1969JChPh..51.2657H} was used as single-particle degrees of freedom.

\subsection{Results and discussion}
As is already well known, the MPS ansatz is able to capture static correlation and hence gives correct potential energy surfaces (PES) whereas HF based methods break down for large interatomic distances.\cite{2006JChPh.125n4101H} The energy per atom as a function of interatomic distance is shown for the 6 spin states in Fig. \ref{energyspinstates}. The energy rises with increasing spin. In the limit of large $R$, all PESs converge in accordance with the non-interacting atom picture.

In the range of $R$ values shown, the equally spaced hydrogen chain is known to make a metal-insulator transition. The transition point is marked by diverging response properties in the TD limit. An earlier ED study has shown that $\alpha_{zz} N^{-2}$ in function of interatomic distance $R$, with $N$ the number of atoms, converges to a limiting curve in the TD limit.\cite{FCIbenda}

The spin dependence of the optical response properties is shown in Fig. \ref{alphaandgammaequally}. For increasing spin, both the polarizability and second hyperpolarizability peaks decrease and shift towards smaller values of $R$. The peaks of the polarizability also occur at slightly smaller values of $R$ than the corresponding peaks of the second hyperpolarizability. Both responses vanish in the limit of large $R$ as a minimal basis set is used.\cite{FCIbenda}

An alternative method to determine the polarizability and second hyperpolarizability is the sum over states (SOS) perturbation expansion.\cite{doi:10.1021/cr00025a008} Note that the dipole moment in the SOS expression commutes with spin operators. Different spin states can hence be treated separately. Two counteracting effects occur in this expression. The number of terms in the summation rapidly decreases with increasing spin because fewer high-spin configurations can be built with $N$ electrons in $L$ orbitals. The magnitude of the terms is expected to be larger for higher spin states due to the smaller energy differences in the denominator. Both effects combined result in properties of the same order of magnitude for the different spin states treated in this paper. The diminishing peak can then be attributed to the smaller number of possible spin configurations. Note that this is only a heuristic argument, as we haven't performed any calculations related to the SOS expression.

\section{A chain of $\text{H}_2$ molecules \label{H2dimerssection}}
In the previous section we have studied a system with changing static correlation. Here we look at a system where the static correlation remains roughly the same, but where the electron delocalization changes.
\subsection{Introduction \label{secConfigDimerExplanation}}
In this section, the optical properties of hydrogen chains with different intra- and intermolecular distances are studied:
\begin{equation}
\vcenter{\hbox{\setlength{\unitlength}{1cm}
\begin{picture}(6.7,1)
\put(0.1,0.375){H}
\put(0.4,0.5){\line(1,0){0.5}}
\put(1.0,0.375){H}
\put(1.3,0.45){......}
\put(1.9,0.375){H}
\put(2.2,0.5){\line(1,0){0.5}}
\put(2.8,0.375){H}
\put(3.1,0.45){......}
\put(3.7,0.375){H}
\put(4.0,0.5){\line(1,0){0.5}}
\put(4.6,0.375){H}
\put(4.9,0.45){......}
\put(5.5,0.375){H}
\put(5.8,0.5){\line(1,0){0.5}}
\put(6.4,0.375){H}
\put(0.5,0.6){$R_f$}
\put(1.4,0.6){$R$}
\put(2.3,0.6){$R_f$}
\put(3.2,0.6){$R$}
\put(4.1,0.6){$R_f$}
\put(5.0,0.6){$R$}
\put(5.9,0.6){$R_f$}
\end{picture}}}
\end{equation}
The intramolecular distance is kept fixed at $R_f = 2$ a.u., whereas the intermolecular distance $R$ can be 2.5, 3 or 4 a.u., in analogy with previous studies.\cite{PhysRevA.52.178, PhysRevA.52.1039, 2009IJQC..109.3103C} In the following, an $\text{H}_2$ constituent will be called a molecule even if $R_f$ is far from the $\text{H}_2$ equilibrium distance. With decreasing $R$, the system changes from a collection of separated $\text{H}_2$ molecules to a chain where the electrons are delocalized,\cite{2009JChPh.131i4104U} whereas the static correlation remains similar due to the constant bond length $R_f$ of the $\text{H}_2$ molecule.

\begin{table}
\begin{tabular}{ c | c  c  c  c  c  c  c }
  $R$ [a.u.] & \multicolumn{7}{c}{$F$ [$10^{-3}$ a.u.]} \\
  \hline
  2.5 & 0.0$~$ & 0.8$~$ & 1.2$~$ & 1.6$~$ & 2.4$~$ & 3.2$~$ & \\
  3.0 & 0.0$~$ & 1.6$~$ & 2.4$~$ & 3.2$~$ & 4.8$~$ & 6.4$~$ & \\
  4.0 & 0.0$~$ & 1.6$~$ & 1.8$~$ & 2.0$~$ & 3.2$~$ & 3.6$~$ & 4.0\\
\end{tabular}

\caption{\label{dentabelFFapplic2} Values of $F$ per intermolecular distance $R$.}
\end{table}

\begin{figure*}[t]
 \includegraphics[width=0.855\textwidth]{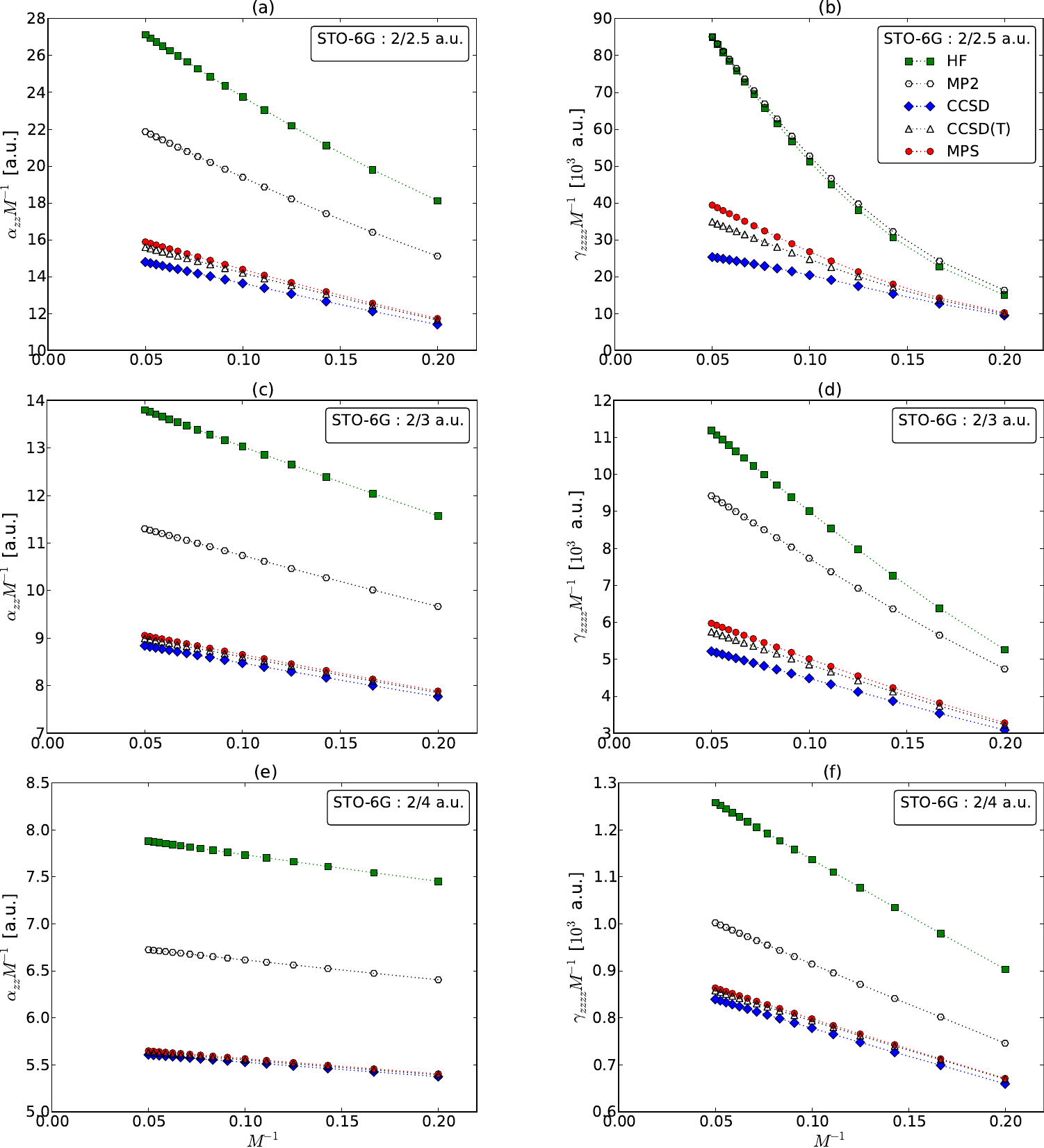}
 \caption{\label{allSTO6G} Polarizabilities and second hyperpolarizabilities of hydrogen chains with intramolecular distance 2 a.u. and varying intermolecular distance, calculated with several LOTs in the L\"owdin transformed STO-6G basis.}
\end{figure*}

Only the absolute ground state ($S=0$) was targeted, but for different chain lengths, levels of theory (LOT) and basis sets. All calculations for the basis sets STO-6G, 6-31G \cite{1972JChPh..56.2257H} and 6-31G(d,p) \cite{citeulike:3179646} were performed with a virtual dimension per symmetry sector $D$ of resp. 32, 64 and 120, independent of chain length and $R$. The fields for which ground state calculations were performed are shown in Table \ref{dentabelFFapplic2}. They depend on the intermolecular distance $R$, but are independent of chain length, basis set and LOT. The LOTs that were studied are MPS, HF, second order M{\o}ller-Plesset perturbation theory (MP2), coupled cluster with singles and doubles (CCSD) and coupled cluster with singles and doubles and perturbative triples (CCSD(T)). The HF, MP2, CCSD and CCSD(T) calculations were performed with the molecular electronic structure program Dalton.\cite{refdalton}

\subsection{Results and discussion}

\begin{figure*}[t]
 \includegraphics[width=0.855\textwidth]{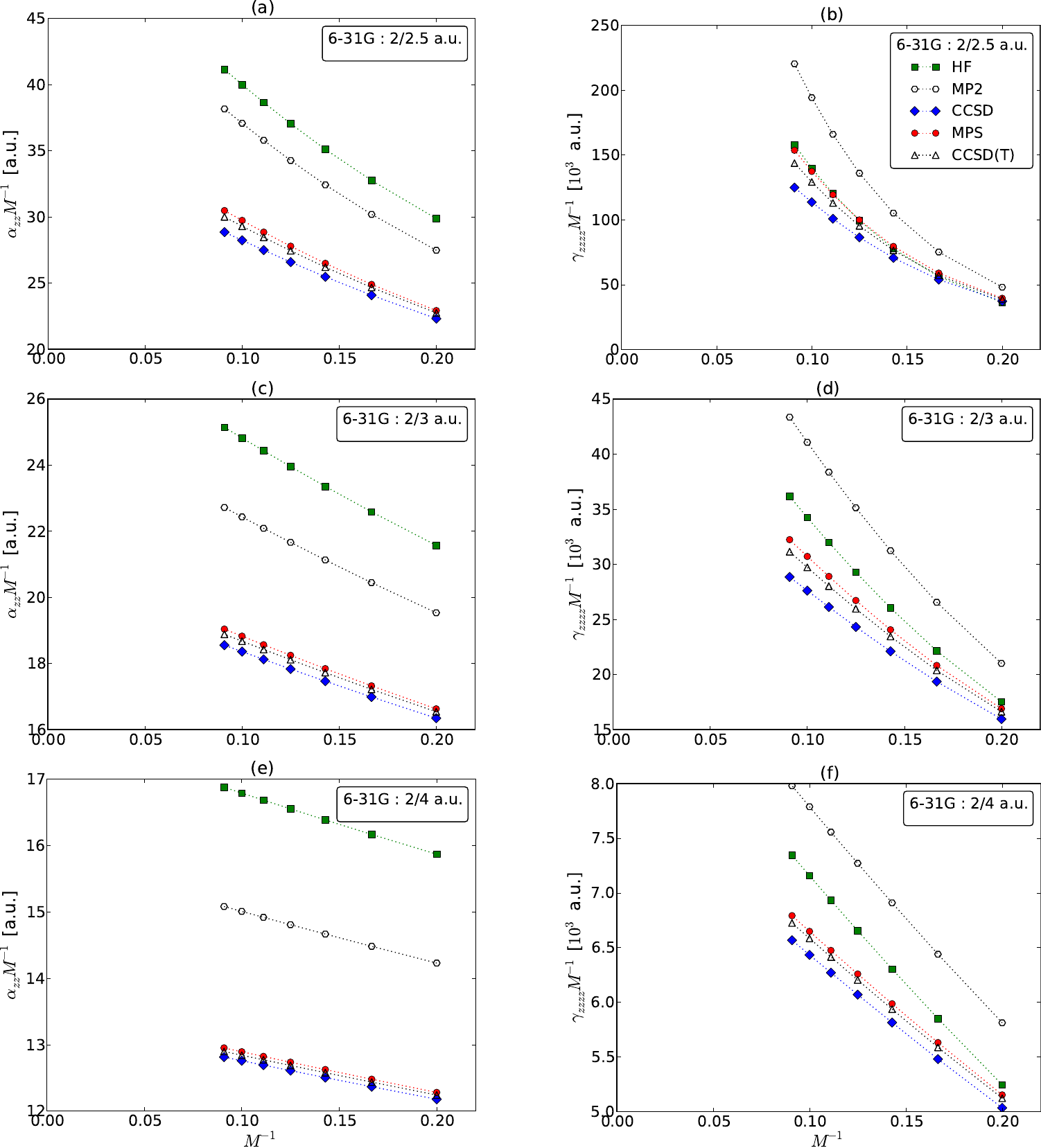}
 \caption{\label{all631G} Polarizabilities and second hyperpolarizabilities of hydrogen chains with intramolecular distance 2 a.u. and varying intermolecular distance, calculated with several LOTs in the L\"owdin transformed 6-31G basis.}
\end{figure*}

For the basis sets STO-6G and 6-31G, $\alpha_{zz}$ and $\gamma_{zzzz}$ were calculated for an increasing number of $\text{H}_2$ units $M$. The values per molecule, $\alpha_{zz}M^{-1}$ and $\gamma_{zzzz}M^{-1}$, are presented for STO-6G in Fig. \ref{allSTO6G} and for 6-31G in Fig. \ref{all631G}, for the different intermolecular distances and LOTs. For the 6-31G(d,p) basis set, the largest chain was $\text{H}_8$. All $\text{H}_8$ data is shown in Table \ref{allH8}.

\begin{table*}
\begin{tabular}{ c | c | l || c  | c  | c | c | c }
  Quantity & R [a.u.] & Basis set & HF & MP2 & CCSD & CCSD(T) & MPS \\
  \hline
  \hline
  \multirow{9}{*}{$\alpha_{zz}$ [a.u.]} & & STO-6G & 63.93 & 53.77 & 41.61 & 42.26 & 42.47\\
  & 2.5 & 6-31G & 105.38 & 96.68 & 80.20 & 81.34 & 81.78\\
  & & 6-31G(d,p) & 106.03 & 102.48 & 91.61 & 92.75 & 93.12\\
  \cline{2-8}
  & & STO-6G & 43.63 & 36.67 & 29.73 & 30.00 & 30.10\\
  & 3.0 & 6-31G & 80.75 & 73.16 & 61.80 & 62.40& 62.66 \\
  & & 6-31G(d,p) & 80.44 & 76.20 & 68.73 & 69.31 & 69.50\\
  \cline{2-8}
  & & STO-6G & 29.26 & 25.21 & 21.20 & 21.27 & 21.31\\
  & 4.0 & 6-31G & 61.77 & 55.46 & 47.62 & 47.84 & 47.97\\
  & & 6-31G(d,p) & 60.90 & 56.49 & 51.52 & 51.70 & 51.77\\
  \hline
  \hline
  \multirow{9}{*}{$\gamma_{zzzz}$ [$10^3$ a.u.]} & & STO-6G & 33.00 & 36.96 & 24.36 & 24.78 & 25.30\\
  & 2.5 & 6-31G & 79.02 & 104.36 & 89.03 & 90.56 & 91.72\\
  & & 6-31G(d,p) & 74.40 & 97.37 & 90.28 & 93.57 & 94.87\\
  \cline{2-8}
  & & STO-6G & 15.50 & 14.33 & 9.90 & 10.20 & 10.30 \\
  & 3.0 & 6-31G & 48.98 & 58.89 & 47.53 & 48.80 & 49.34\\
  & & 6-31G(d,p) & 47.25 & 58.10 & 49.78 & 51.98 & 52.62\\
  \cline{2-8}
  & & STO-6G & 3.17 & 2.66 & 2.41 & 2.44 & 2.44\\
  & 4.0 & 6-31G & 17.63 & 19.75 & 17.60 & 17.85 & 17.92\\
  & & 6-31G(d,p) & 17.38 & 19.53 & 17.42 & 17.88 & 18.00\\
  \hline
  \hline
\end{tabular}

\caption{\label{allH8} All polarizability and second hyperpolarizability data for $\text{H}_8$.}
\end{table*}

From Table \ref{allH8}, it can be observed that for corresponding intermolecular distances and LOTs, the STO-6G polarizability and second hyperpolarizability values are significantly lower than the values obtained with the 6-31G and 6-31G(d,p) basis sets. The possible movement of electrons in a minimal basis set is of course restricted. The 6-31G and 6-31G(d,p) results are also much closer to each other than to the minimal basis set results, in agreement with Champagne \textit{et al.}\cite{PhysRevA.52.178, PhysRevA.52.1039}

For the polarizability of long chains, a clear order exists for the LOTs, which is the same for the three intermolecular distances and the STO-6G and 6-31G basis sets:
\begin{equation}
\alpha_{zz}^{HF} > \alpha_{zz}^{MP2} > \alpha_{zz}^{MPS} > \alpha_{zz}^{CCSD(T)} > \alpha_{zz}^{CCSD} \label{polorder}
\end{equation}
This order is in agreement with previous work,\cite{PhysRevA.52.1039} which looks at small basis sets. For larger basis sets, it was found that the HF polarizability tends to drop below the MP2 values for decreasing values of the intermolecular distance $R$ (increasing electron delocalization).\cite{2009IJQC..109.3103C} There is also a clear order in the deviation between the polarizability obtained with a certain LOT and the MPS result:
\begin{equation}
\Delta \alpha_{zz}^{HF} > \Delta \alpha_{zz}^{MP2} > \Delta \alpha_{zz}^{CCSD} > \Delta \alpha_{zz}^{CCSD(T)} \label{hyperpolorder}
\end{equation}

\begin{figure*}[t]
 \includegraphics[width=0.855\textwidth]{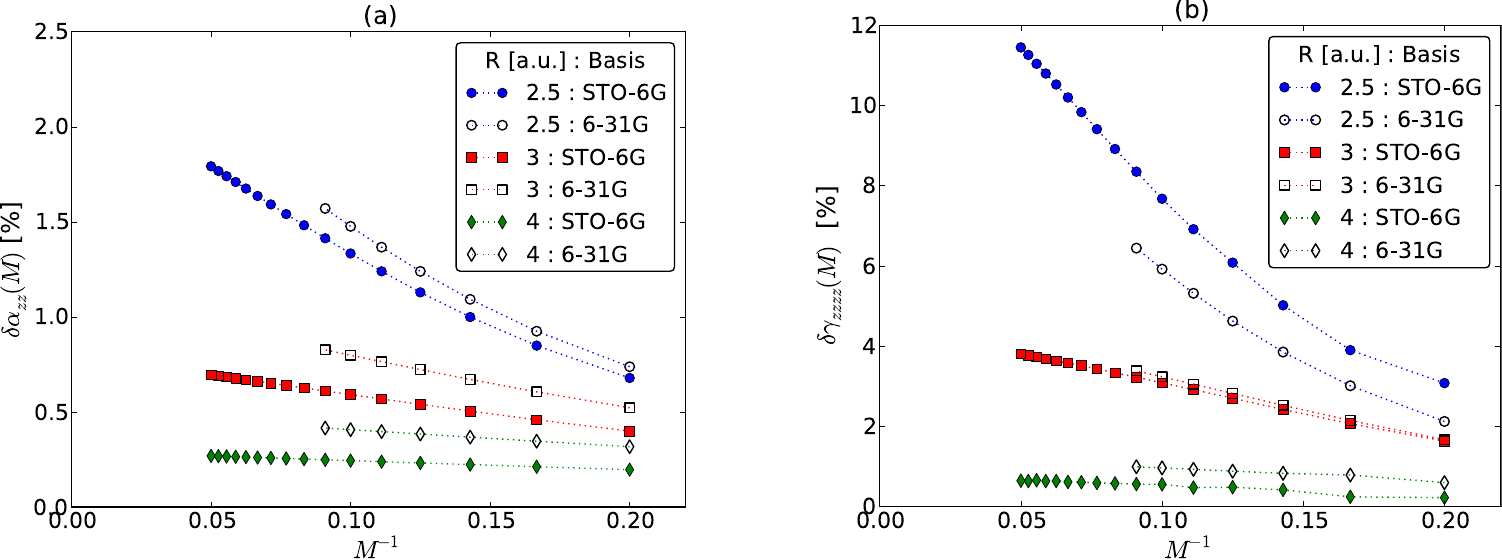}
 \caption{\label{reldeviationplot} The relative deviation (see Eq. \eqref{reldeveq}) of the CCSD(T) polarizability (a) and second hyperpolarizability (b) to the MPS values for the data in Fig. \ref{allSTO6G} and Fig. \ref{all631G}.}
\end{figure*}

For the second hyperpolarizability of long chains, a clear order exists for all LOTs except HF. Again this order is the same for the three intermolecular distances and the STO-6G and 6-31G basis sets, and equals the one in \eqref{polorder} when $\alpha_{zz}^{HF}$ is excluded. The HF second hyperpolarizability tends to drop below the MP2 values for decreasing values of the intermolecular distance $R$ (increasing electron delocalization) and for increasing basis sets. For even larger basis sets, the HF values drop below the CCSD values, but the order of the other methods is also left unchanged.\cite{2009IJQC..109.3103C} It is intriguing that for the second hyperpolarizability, the mean-field (HF) results have no fixed position relative to the other correlated methods. This shows that the approximate treatment of electron correlation by MP2 or CCSD and CCSD(T) does not lead to a smooth transition from mean-field theory towards ED. Instead, the final value of $\gamma_{zzzz}$ is the result of a delicate balance of positive and negative contributions from the various excited determinants that are summed up with different weights. This fluctuating nature of electron correlation on NLO properties was also observed in linearly $\pi$ conjugated chains.\cite{2011JChPh.135a4111L} For the deviations, the same order as in \eqref{hyperpolorder} is found, when $\Delta \alpha_{zz}^{HF}$ is excluded.

CCSD(T) is often used as the benchmark method to test the performance of LOTs for linear and non-linear optical properties.\cite{2009IJQC..109.3103C} Of the four HF based LOTs we have tested, CCSD(T) indeed consistently gives the best results. To check the performance of CCSD(T) for the data in Fig. \ref{allSTO6G} and Fig. \ref{all631G}, the relative deviation
\begin{equation}
 \delta q (M) = \frac{q^{MPS}(M) - q^{CCSD(T)}(M)}{q^{MPS}(M)} \label{reldeveq}
\end{equation}
is defined. $q$ can again be $\alpha_{zz}$ or $\gamma_{zzzz}$. This relative deviation is shown in Fig. \ref{reldeviationplot}. The deviation is larger for the second hyperpolarizability than for the polarizability. For both parameters, the deviation increases with decreasing intermolecular distance (increasing electron delocalization). For chains with small intermolecular distance (delocalized electrons), the deviation also rapidly increases with the number of molecules. Note that the $\gamma_{zzzz}^{CCSD(T)}(M=20)$ result for the intermolecular distance $R$ = 2.5 a.u. and the STO-6G basis set already deviates by 12\% from the exact result and a simple extrapolation to the TD limit shows that this deviation can become as large as 15\%. The breakdown of the CCSD(T) method can be understood by the following heuristic argument in terms of elementary optical excitations. For large intermolecular distances, the electrons are localized in $\text{H}_2$ molecules and the maximum number of electrons involved in an elementary excitation is 2. These effects can be captured by the CCSD(T) method. For small intermolecular distances, the electrons are delocalized over the chain and a larger number of electrons are involved in elementary excitations. This number also increases with chain length. CCSD(T) cannot adequately capture this effect and the CCSD(T) results start to deviate from the exact ones.

\begin{figure}
\includegraphics[width=0.40\textwidth]{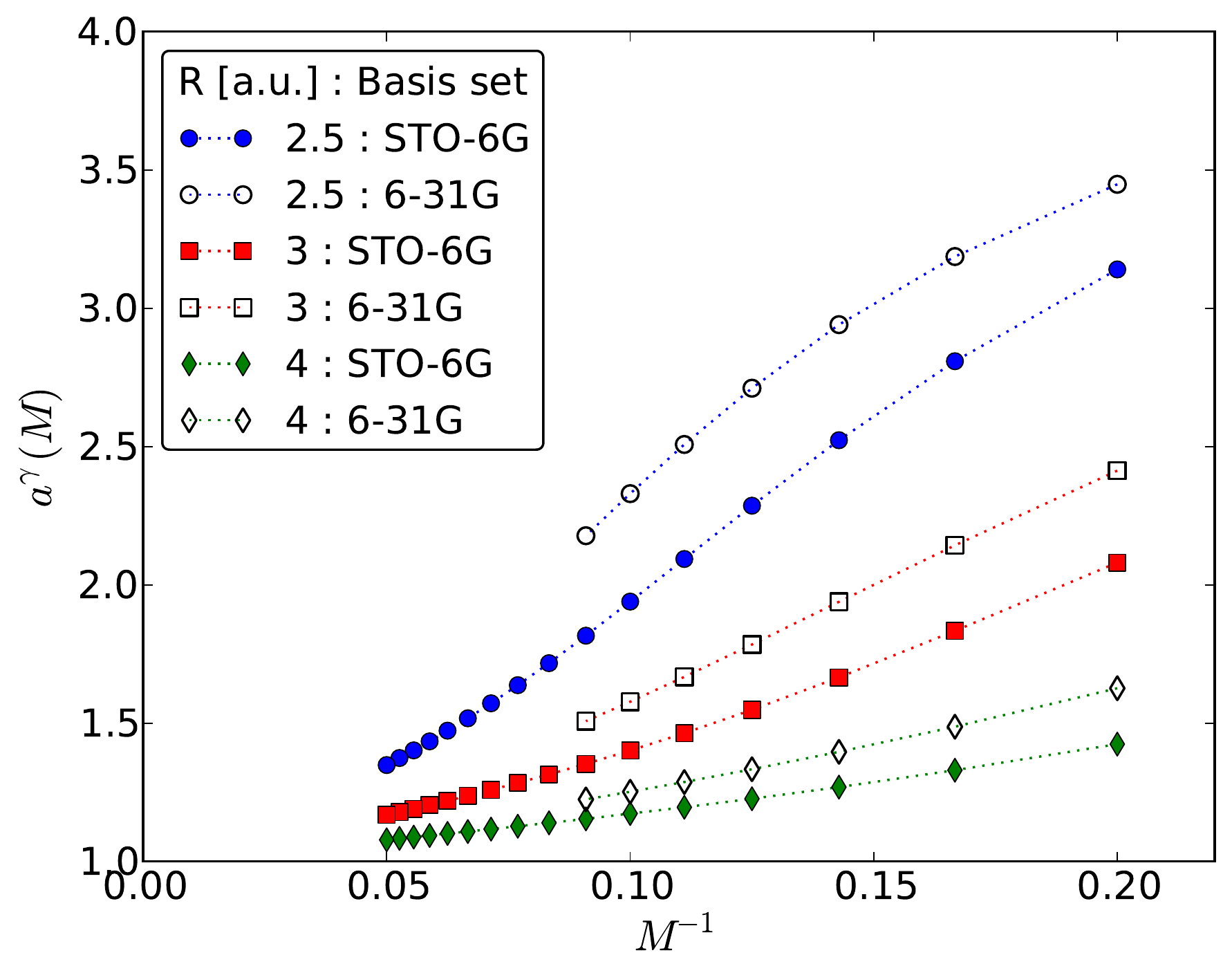}
\caption{\label{powergamma} The power approximation of Eq. \eqref{powerapproxeq}, applied to the MPS calculations for the STO-6G and 6-31G basis sets.}
\end{figure}

For the second hyperpolarizability, the scaling
\begin{equation}
 \gamma(M) \propto M^{a(M)}
\end{equation}
is often proposed.\cite{doi:10.1021/cr00025a008} The power $a(M)$ depends weakly on the number of molecules $M$. Its initially constant value drops eventually towards one in the TD limit. This can be explained in terms of a delocalized optical excitation, with a typical length scale. With increasing lengths, the possibility for such excitations opens up. When the chain can contain the delocalized excitations completely, the power tends to 1 and it is said that the system is in the saturation regime.\cite{doi:10.1021/cr00025a008} As can be seen in Fig. \ref{allSTO6G} and Fig. \ref{all631G}, the saturation regime indeed sets in later when the intermolecular distance is smaller (electron delocalization larger). This can be confirmed by the following approximation to $a(M)$:
\begin{equation}
a^{\gamma}(M) = \frac{\ln{ (\gamma_{zzzz}(M) )} - \ln{(\gamma_{zzzz}(M-1))} }{\ln{(M)} - \ln{(M-1})} \label{powerapproxeq}
\end{equation}
which is shown in Fig. \ref{powergamma} for the MPS calculations. From this figure, two extra conclusions can be made. The power for $R$ = 2.5 a.u. and the 6-31G basis set is still above 2 for the chain lengths studied. Accurate extrapolations of the second hyperpolarizability to the TD limit are therefore not possible for this data set. The estimated powers are larger for the 6-31G basis than for the STO-6G basis, a result of the increased number of possibilities for optical excitations in 6-31G, but the effect of electron delocalization predominates.

\begin{figure}
 \includegraphics[width=0.40\textwidth]{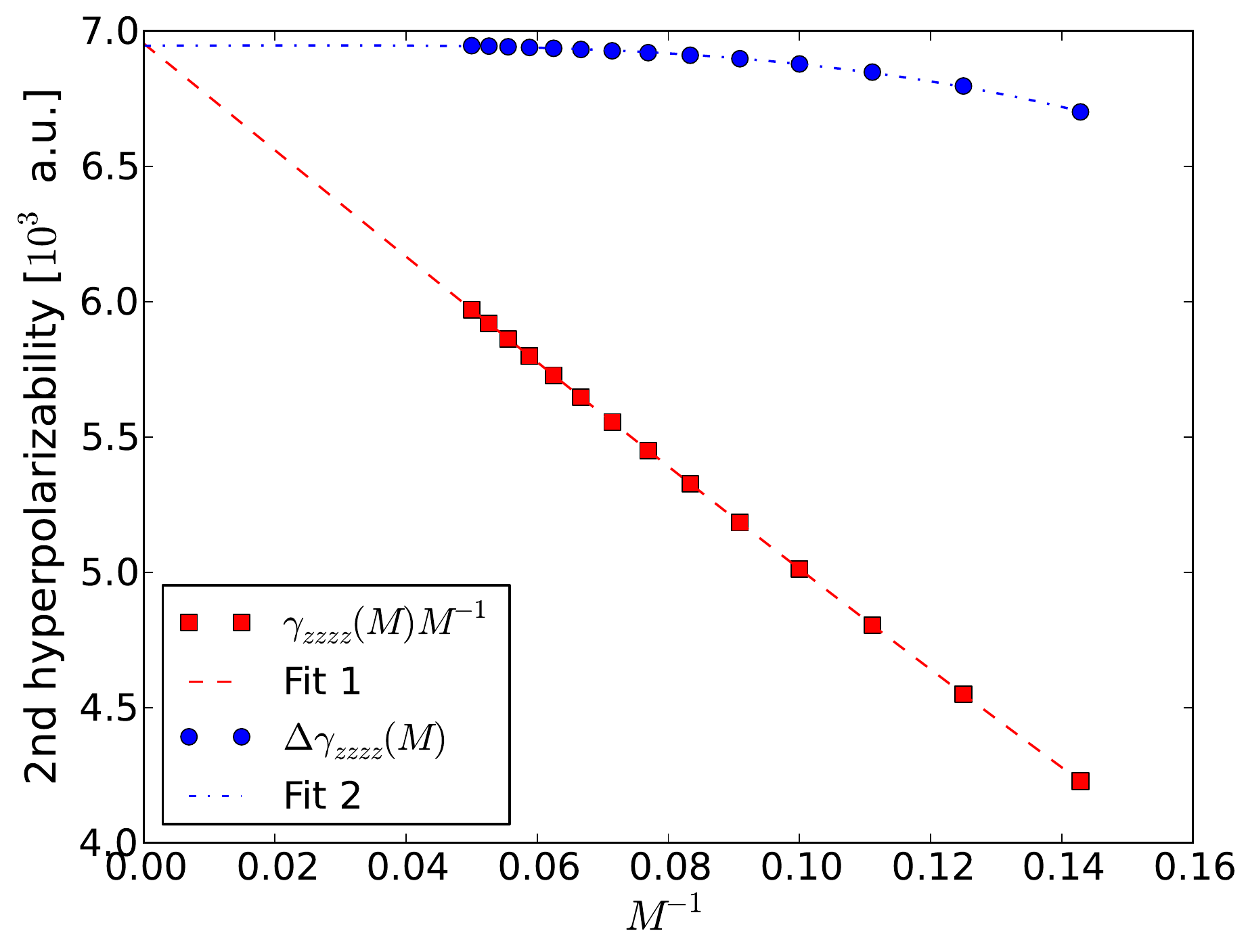}
 \caption{\label{extrapolexamplefig} Extrapolation of the second hyperpolarizability per $\text{H}_2$ unit for the configuration with $R$ = 3.0 a.u. in the STO-6G basis. The extrapolation schemes in Eq. \eqref{propfit} and Eq. \eqref{propfit2} were used to obtain resp. Fit 1 and 2.}
\end{figure}

From the data in Fig. \ref{allSTO6G} and Fig. \ref{all631G}, values for $\alpha_{zz}^{MPS} M^{-1}$ and $\gamma_{zzzz}^{MPS} M^{-1}$ in the TD limit can be extrapolated. A scaling relation of the form
\begin{equation}
\frac{q(M)}{M} = a_0 + \frac{a_1}{M} + \frac{a_2}{M^2} + \frac{a_3}{M^3} \label{propfit}
\end{equation}
is assumed, where $q$ can again be $\alpha_{zz}$ or $\gamma_{zzzz}$ and the $a_n$ are obtained from a least squares fit. The parameter $a_0$ then corresponds to the desired TD limit value. From \eqref{propfit} the following equation can be derived:
\begin{eqnarray}
\Delta q(M) & = & q(M) - q(M-1) \nonumber\\
& = & a_0 + \frac{b_2}{M^2} + \frac{b_3}{M^3} + \mathcal{O}(M^{-4}) \label{propfit2}
\end{eqnarray}
To check the extrapolations, a least squares fit of \eqref{propfit2} to $\Delta q(M)$ is performed too. In both extrapolation schemes the cut-off value for $M$ was 5 for the polarizability and 7 for the second hyperpolarizability. An example is shown in Fig. \ref{extrapolexamplefig}. All obtained data are presented in Table \ref{extrapoltable}. Except for the second hyperpolarizability for $R$ = 2.5 a.u. and the 6-31G basis, the results of both extrapolation schemes are within 1\% relative deviation.

\begin{table}
\begin{tabular}{ c | l | c || c  | c }
  Quantity & Basis set & R [a.u.] & Eq. \eqref{propfit} & Eq. \eqref{propfit2}\\
  \hline
  \hline
  \multirow{6}{*}{$\alpha_{zz}$ [a.u.]} & & 2.5 & 17.41 & 17.41 \\
  & STO-6G & 3.0 & 9.464 & 9.462  \\
  & & 4.0 & 5.733 & 5.733 \\
  \cline{2-5}
  & & 2.5 & 39.00 & 39.20 \\
  & 6-31G & 3.0 & 21.27 & 21.27 \\
  & & 4.0 & 13.55 & 13.55\\
  \hline
  \hline
  \multirow{6}{*}{$\gamma_{zzzz}$ [$10^3$ a.u.]} & & 2.5 & 52.64 & 52.74\\
  & STO-6G & 3.0 & 6.953 & 6.945\\
  & & 4.0 & 0.9303 & 0.9301 \\
  \cline{2-5}
  & & 2.5 & 410.8$^{(a)}$ & 424.0$^{(a)}$\\
  & 6-31G & 3.0 & 48.52 & 48.45 \\
  & & 4.0 & 8.275 & 8.269 \\
  \hline
  \hline
\end{tabular}

\caption{\label{extrapoltable} Extrapolated values for the polarizability and second hyperpolarizability per $\text{H}_2$ unit in the TD limit. (a) These extrapolated values lie far apart and have to be treated with care as the powers $a^{\gamma}(M)$ for the largest chain lengths studied are still rather large.}
\end{table}

\section{Conclusions \label{sectconclusions}}
There is a lot of interest in the optical properties of chemical systems extended in one spatial dimension. The MPS ansatz works well for quasi-one-dimensional non-critical systems and yields highly accurate results. It can hence be used to study the optical properties of one-dimensional systems. We have implemented the sweep algorithm for the variational optimization of SU(2) $\otimes$ U(1) invariant MPSs to study the static longitudinal polarizability and second hyperpolarizability of hydrogen chains by means of finite field extrapolations.

As a first application, the optical response properties of an equally spaced hydrogen chain were studied for the ground states in different spin symmetry sectors. It is well known that HF based methods break down in the limit of large interatomic distances, whereas an MPS can capture the relevant static correlation needed to obtain accurate energy results. It was shown that accurate optical response properties can also be obtained with the MPS ansatz. The peaks of the polarizability and second hyperpolarizability decrease with increasing spin and shift towards smaller interatomic distances. Arguments based on a SOS expansion can be invoked to explain which terms contribute to these optical response properties.

CCSD(T) is often used as a reference method for the calculation of optical response properties. For roughly constant static correlation, avoiding the expected breakdown of HF based methods, the deviation of the optical properties calculated with CCSD(T) and the quasi-exact MPS method was studied. For increasing electron delocalization, the deviation becomes larger. For a large electron delocalization, the deviation rapidly increases with increasing chain length. The increasing deviation was explained in terms of delocalized optical excitations, which CCSD(T) cannot accurately capture. For small basis sets, the MPS algorithm gives accurate optical response properties in the saturation regime. These results were extrapolated to the TD limit.

In the future, we aim to implement the quadratically scaling algorithm of Hachmann \textit{et al.} \cite{2006JChPh.125n4101H} and try to find a better choice of virtual dimension truncation to extend the range of our algorithm. We also aim to extend our algorithm to find excited states, allowing a study of the dominant terms in the SOS expression.
 
The MPS algorithm is hence a promising method to assess the performance of other QC methods for quasi-one-dimensional chemical systems. It allows to maintain ED accuracy for larger system sizes, e.g. to obtain accurate results of optical response properties in the saturation regime.

\begin{acknowledgements}
This research was supported by the Research Foundation Flanders (S.W.), the Swiss National Science Foundation fellowship PBEZP2-134449 (P.A.L.), and NSERC (P.W.A.). The authors acknowledge a generous allocation of computer time granted by the Stevin Supercomputer Infrastructure at Ghent University, funded by Ghent University, the Hercules Foundation and the Flemish Government - department EWI. Additional computing ressources were granted by SHARCNET, a partner consortium in the Compute Canada national HPC platform.
\end{acknowledgements}

\section*{Appendix A}
Note that during a sweep, we work with left normalized tensors in the left part and right normalized tensors in the right part. Consider the following partial contraction in the graphical notation:\cite{schollwockDMRGatMPSage}
\begin{equation}
\vcenter{\hbox{\scriptsize{
\setlength{\unitlength}{1cm}
\begin{picture}(2.6,2)
\put(0.4,1.8){\circle{0.4}}
\put(0.6,1.8){\line(1,0){0.75}}
\put(1.4,1.7){$j_R j_R^z N_R \alpha_R$}
\put(0.28,1.7){\scriptsize{M}}
\put(0.4,1.3){\line(0,1){0.3}}
\put(0.1,0.7){\line(1,0){0.6}}
\put(0.1,1.3){\line(1,0){0.6}}
\put(0.1,0.7){\line(0,1){0.6}}
\put(0.7,0.7){\line(0,1){0.6}}
\put(0.2,0.9){$a_m$}
\put(0.4,0.4){\line(0,1){0.3}}
\put(0.4,0.2){\circle{0.4}}
\put(0.6,0.2){\line(1,0){0.75}}
\put(0.28,0.1){M}
\put(1.40,0.1){$\widetilde{j}_R \widetilde{j}_R^z \widetilde{N}_R \widetilde{\alpha}_R$}
\put(0.2,1.0){\oval(1.0,1.6)[l]}
\end{picture}
}}} \label{example1}
\end{equation}
\begin{widetext}
With \eqref{tensordecomp}, it is easy to show that \eqref{example1} can be written as
\begin{equation}
\delta_{\widetilde{N}_R,N_R+1} \braket{j_R j_R^z \frac{1}{2} m \mid \widetilde{j}_R \widetilde{j}_R^z} \vcenter{\hbox{\scriptsize{
\setlength{\unitlength}{1cm}
\begin{picture}(2.0,2.4)
\put(0.2,0.9){\line(1,0){0.4}}
\put(0.2,1.5){\line(1,0){0.4}}
\put(0.2,0.9){\line(0,1){0.6}}
\put(0.6,0.9){\line(0,1){0.6}}
\put(0.3,1.1){$\Lambda$}
\put(0.4,1.5){\line(0,1){0.5}}
\put(0.4,0.4){\line(0,1){0.5}}
\put(0.4,2.0){\line(1,0){0.75}}
\put(0.4,0.4){\line(1,0){0.75}}
\put(1.2,1.9){$j_R N_R \alpha_R$}
\put(1.2,0.3){$\widetilde{j}_R (N_R + 1) \widetilde{\alpha}_R$}
\end{picture}
}}} \label{againWE}
\vspace{0.01\textwidth}
\end{equation}
with
\begin{equation}
\vcenter{\hbox{\scriptsize{
\setlength{\unitlength}{1cm}
\begin{picture}(2.8,2.4)
\put(0.2,0.9){\line(1,0){0.4}}
\put(0.2,1.5){\line(1,0){0.4}}
\put(0.2,0.9){\line(0,1){0.6}}
\put(0.6,0.9){\line(0,1){0.6}}
\put(0.3,1.1){$\Lambda$}
\put(0.4,1.5){\line(0,1){0.5}}
\put(0.4,0.4){\line(0,1){0.5}}
\put(0.4,2.0){\line(1,0){0.75}}
\put(0.4,0.4){\line(1,0){0.75}}
\put(1.2,1.9){$j_R N_R \alpha_R$}
\put(1.2,0.3){$\widetilde{j}_R (N_R + 1) \widetilde{\alpha}_R$}
\end{picture}
}}}
= \sum\limits_{\alpha_L} \vcenter{\hbox{\scriptsize{
\setlength{\unitlength}{1cm}
\begin{picture}(3.7,2)
\put(1.1,1.8){\circle{0.4}}
\put(1.3,1.8){\line(1,0){0.75}}
\put(2.1,1.7){$j_R N_R \alpha_R$}
\put(1.0,1.7){T}
\put(1.1,1.2){\line(0,1){0.4}}
\put(1.2,1.3){$0 0$}
\put(1.1,0.4){\line(0,1){0.4}}
\put(1.2,0.5){$\frac{1}{2} 1$}
\put(1.1,0.2){\circle{0.4}}
\put(1.3,0.2){\line(1,0){0.75}}
\put(1.0,0.1){T}
\put(2.1,0.1){$\widetilde{j}_R (N_R+1) \widetilde{\alpha}_R$}
\put(0.9,1.0){\oval(1.0,1.6)[l]}
\put(0.0,0.4){\rotatebox{90}{$j_R N_R \alpha_L$}}
\end{picture}
}}}
+ (-1)^{\widetilde{j}_R - j_R + \frac{1}{2}} \sqrt{\frac{2 j_R + 1}{2 \widetilde{j}_R + 1}}\sum\limits_{\alpha_L} \vcenter{\hbox{\scriptsize{
\setlength{\unitlength}{1cm}
\begin{picture}(3.7,2)
\put(1.1,1.8){\circle{0.4}}
\put(1.3,1.8){\line(1,0){0.75}}
\put(2.1,1.7){$j_R N_R \alpha_R$}
\put(1.0,1.7){T}
\put(1.1,1.2){\line(0,1){0.4}}
\put(1.2,1.3){$\frac{1}{2} 1$}
\put(1.1,0.4){\line(0,1){0.4}}
\put(1.2,0.5){$0 2$}
\put(1.1,0.2){\circle{0.4}}
\put(1.3,0.2){\line(1,0){0.75}}
\put(1.0,0.1){T}
\put(2.1,0.1){$\widetilde{j}_R (N_R+1) \widetilde{\alpha}_R$}
\put(0.9,1.0){\oval(1.0,1.6)[l]}
\put(0.0,0.0){\rotatebox{90}{$\widetilde{j}_R (N_R-1) \alpha_L$}}
\end{picture}
}}}
\end{equation}
Eq. \eqref{example1} can hence be decomposed into a structural part (Clebsch-Gordan coefficient and particle conserving Kronecker delta) and a degeneracy part (the reduced $\Lambda$ tensor with spin $\frac{1}{2}$), as is shown in \eqref{againWE}. As a second example, consider the partial contraction:
\begin{align}
\vcenter{\hbox{\scriptsize{
\setlength{\unitlength}{1cm}
\begin{picture}(4.0,2)
\put(0.4,1.8){\circle{0.4}}
\put(1.6,1.8){\line(1,0){0.75}}
\put(2.4,1.7){$j_R j_R^z N_R \alpha_R$}
\put(0.28,1.7){\scriptsize{M}}
\put(0.4,1.3){\line(0,1){0.3}}
\put(0.1,0.7){\line(1,0){0.6}}
\put(0.1,1.3){\line(1,0){0.6}}
\put(0.1,0.7){\line(0,1){0.6}}
\put(0.7,0.7){\line(0,1){0.6}}
\put(0.19,0.9){$a_{m_1}$}
\put(0.4,0.4){\line(0,1){0.3}}
\put(0.4,0.2){\circle{0.4}}
\put(0.6,1.8){\line(1,0){0.6}}
\put(0.6,0.2){\line(1,0){0.6}}
\put(1.4,1.8){\circle{0.4}}
\put(1.4,0.2){\circle{0.4}}
\put(1.1,0.7){\line(0,1){0.6}}
\put(1.7,0.7){\line(0,1){0.6}}
\put(1.1,0.7){\line(1,0){0.6}}
\put(1.1,1.3){\line(1,0){0.6}}
\put(1.4,1.3){\line(0,1){0.3}}
\put(1.19,0.9){$a_{m_2}^{\dagger}$}
\put(1.4,0.4){\line(0,1){0.3}}
\put(1.28,1.7){\scriptsize{M}}
\put(1.28,0.1){\scriptsize{M}}
\put(1.6,0.2){\line(1,0){0.75}}
\put(0.28,0.1){M}
\put(2.40,0.1){$\widetilde{j}_R \widetilde{j}_R^z \widetilde{N}_R \widetilde{\alpha}_R$}
\put(0.2,1.0){\oval(1.0,1.6)[l]}
\end{picture}
}}} = & \delta_{N_R,\widetilde{N}_R} (-1)^{\frac{1}{2}-m_2} \left( \braket{\frac{1}{2} m_1 \frac{1}{2} -m_2 \mid 0 0} \braket{j_R j_R^z 0 0 \mid \widetilde{j}_R \widetilde{j}_R^z} \vcenter{\hbox{\scriptsize{
\setlength{\unitlength}{1cm}
\begin{picture}(2.0,2.4)
\put(0.1,0.9){\line(1,0){0.6}}
\put(0.1,1.5){\line(1,0){0.6}}
\put(0.1,0.9){\line(0,1){0.6}}
\put(0.7,0.9){\line(0,1){0.6}}
\put(0.2,1.1){$F^0$}
\put(0.4,1.5){\line(0,1){0.5}}
\put(0.4,0.4){\line(0,1){0.5}}
\put(0.4,2.0){\line(1,0){0.75}}
\put(0.4,0.4){\line(1,0){0.75}}
\put(1.2,1.9){$j_R N_R \alpha_R$}
\put(1.2,0.3){$j_R N_R \widetilde{\alpha}_R$}
\end{picture}
}}} \right. \nonumber \\ 
+ & \left. \braket{\frac{1}{2} m_1 \frac{1}{2} -m_2 \mid 1 (m_1 - m_2)} \braket{j_R j_R^z 1 (m_1-m_2) \mid \widetilde{j}_R \widetilde{j}_R^z} \vcenter{\hbox{\scriptsize{
\setlength{\unitlength}{1cm}
\begin{picture}(2.0,2.4)
\put(0.1,0.9){\line(1,0){0.6}}
\put(0.1,1.5){\line(1,0){0.6}}
\put(0.1,0.9){\line(0,1){0.6}}
\put(0.7,0.9){\line(0,1){0.6}}
\put(0.2,1.1){$F^1$}
\put(0.4,1.5){\line(0,1){0.5}}
\put(0.4,0.4){\line(0,1){0.5}}
\put(0.4,2.0){\line(1,0){0.75}}
\put(0.4,0.4){\line(1,0){0.75}}
\put(1.2,1.9){$j_R N_R \alpha_R$}
\put(1.2,0.3){$\widetilde{j}_R N_R \widetilde{\alpha}_R$}
\end{picture}
}}} \quad \right)
\end{align}
with
\begin{align}
\vcenter{\hbox{\scriptsize{
\setlength{\unitlength}{1cm}
\begin{picture}(2.8,2.4)
\put(0.1,0.9){\line(1,0){0.6}}
\put(0.1,1.5){\line(1,0){0.6}}
\put(0.1,0.9){\line(0,1){0.6}}
\put(0.7,0.9){\line(0,1){0.6}}
\put(0.2,1.1){$F^0$}
\put(0.4,1.5){\line(0,1){0.5}}
\put(0.4,0.4){\line(0,1){0.5}}
\put(0.4,2.0){\line(1,0){0.75}}
\put(0.4,0.4){\line(1,0){0.75}}
\put(1.2,1.9){$j_R N_R \alpha_R$}
\put(1.2,0.3){$j_R N_R \widetilde{\alpha}_R$}
\end{picture}
}}}
= & \sum\limits_{j_L \alpha_L \widetilde{\alpha}_L} \frac{1}{\sqrt{2}} \vcenter{\hbox{\scriptsize{
\setlength{\unitlength}{1cm}
\begin{picture}(4.4,2.4)
\put(0.6,0.9){\line(1,0){0.4}}
\put(0.6,1.5){\line(1,0){0.4}}
\put(0.6,0.9){\line(0,1){0.6}}
\put(1.0,0.9){\line(0,1){0.6}}
\put(0.7,1.1){$\Lambda$}
\put(0.8,1.5){\line(0,1){0.5}}
\put(0.8,0.4){\line(0,1){0.5}}
\put(0.8,2.0){\line(1,0){1.0}}
\put(0.8,0.4){\line(1,0){1.0}}
\put(2.0,2.0){\circle{0.4}}
\put(2.0,0.4){\circle{0.4}}
\put(1.9,1.9){T}
\put(1.9,0.3){T}
\put(2.0,1.4){\line(0,1){0.4}}
\put(2.0,0.6){\line(0,1){0.4}}
\put(2.1,1.5){$\frac{1}{2} 1$}
\put(2.1,0.7){$0 0$}
\put(2.2,0.4){\line(1,0){0.4}}
\put(2.2,2.0){\line(1,0){0.4}}
\put(2.7,1.9){$j_R N_R \alpha_R$}
\put(2.7,0.3){$j_R N_R \widetilde{\alpha}_R$}
\put(0.0,2.1){$j_L (N_R-1) \alpha_L$}
\put(0.8,0.1){$j_R N_R \widetilde{\alpha}_L$}
\end{picture}
}}} \nonumber \\
+ & \sum\limits_{\widetilde{j}_L \alpha_L \widetilde{\alpha}_L} \frac{1}{\sqrt{2}} \sqrt{\frac{2 \widetilde{j}_L+1}{2 j_R+1}} (-1)^{j_R - \widetilde{j}_L + \frac{1}{2}} \vcenter{\hbox{\scriptsize{
\setlength{\unitlength}{1cm}
\begin{picture}(4.4,2.4)
\put(0.6,0.9){\line(1,0){0.4}}
\put(0.6,1.5){\line(1,0){0.4}}
\put(0.6,0.9){\line(0,1){0.6}}
\put(1.0,0.9){\line(0,1){0.6}}
\put(0.7,1.1){$\Lambda$}
\put(0.8,1.5){\line(0,1){0.5}}
\put(0.8,0.4){\line(0,1){0.5}}
\put(0.8,2.0){\line(1,0){1.0}}
\put(0.8,0.4){\line(1,0){1.0}}
\put(2.0,2.0){\circle{0.4}}
\put(2.0,0.4){\circle{0.4}}
\put(1.9,1.9){T}
\put(1.9,0.3){T}
\put(2.0,1.4){\line(0,1){0.4}}
\put(2.0,0.6){\line(0,1){0.4}}
\put(2.1,1.5){$0 2$}
\put(2.1,0.7){$\frac{1}{2} 1$}
\put(2.2,0.4){\line(1,0){0.4}}
\put(2.2,2.0){\line(1,0){0.4}}
\put(2.7,1.9){$j_R N_R \alpha_R$}
\put(2.7,0.3){$j_R N_R \widetilde{\alpha}_R$}
\put(0.0,2.1){$j_R (N_R-2) \alpha_L$}
\put(0.0,0.1){$\widetilde{j}_L (N_R - 1) \widetilde{\alpha}_L$}
\end{picture}
}}}
\end{align}
and
\begin{align}
\vcenter{\hbox{\scriptsize{
\setlength{\unitlength}{1cm}
\begin{picture}(2.8,2.4)
\put(0.1,0.9){\line(1,0){0.6}}
\put(0.1,1.5){\line(1,0){0.6}}
\put(0.1,0.9){\line(0,1){0.6}}
\put(0.7,0.9){\line(0,1){0.6}}
\put(0.2,1.1){$F^1$}
\put(0.4,1.5){\line(0,1){0.5}}
\put(0.4,0.4){\line(0,1){0.5}}
\put(0.4,2.0){\line(1,0){0.75}}
\put(0.4,0.4){\line(1,0){0.75}}
\put(1.2,1.9){$j_R N_R \alpha_R$}
\put(1.2,0.3){$\widetilde{j}_R N_R \widetilde{\alpha}_R$}
\end{picture}
}}}
= & \sum\limits_{j_L \alpha_L \widetilde{\alpha}_L} \sqrt{3(2 j_R + 1)} (-1)^{\widetilde{j}_R + j_L + \frac{3}{2}} \left\{ \begin{array}{ccc} \frac{1}{2} & \frac{1}{2} & 1 \\ j_R & \widetilde{j}_R & j_L \end{array} \right\} \vcenter{\hbox{\scriptsize{
\setlength{\unitlength}{1cm}
\begin{picture}(4.4,2.4)
\put(0.6,0.9){\line(1,0){0.4}}
\put(0.6,1.5){\line(1,0){0.4}}
\put(0.6,0.9){\line(0,1){0.6}}
\put(1.0,0.9){\line(0,1){0.6}}
\put(0.7,1.1){$\Lambda$}
\put(0.8,1.5){\line(0,1){0.5}}
\put(0.8,0.4){\line(0,1){0.5}}
\put(0.8,2.0){\line(1,0){1.0}}
\put(0.8,0.4){\line(1,0){1.0}}
\put(2.0,2.0){\circle{0.4}}
\put(2.0,0.4){\circle{0.4}}
\put(1.9,1.9){T}
\put(1.9,0.3){T}
\put(2.0,1.4){\line(0,1){0.4}}
\put(2.0,0.6){\line(0,1){0.4}}
\put(2.1,1.5){$\frac{1}{2} 1$}
\put(2.1,0.7){$0 0$}
\put(2.2,0.4){\line(1,0){0.4}}
\put(2.2,2.0){\line(1,0){0.4}}
\put(2.7,1.9){$j_R N_R \alpha_R$}
\put(2.7,0.3){$\widetilde{j}_R N_R \widetilde{\alpha}_R$}
\put(0.0,2.1){$j_L (N_R-1) \alpha_L$}
\put(0.8,0.1){$\widetilde{j}_R N_R \widetilde{\alpha}_L$}
\end{picture}
}}} \nonumber \\
+ & \sum\limits_{\widetilde{j}_L \alpha_L \widetilde{\alpha}_L} \sqrt{3(2 \widetilde{j}_L + 1)} (-1)^{\widetilde{j}_R + j_R + 1} \left\{ \begin{array}{ccc} \frac{1}{2} & \frac{1}{2} & 1 \\ j_R & \widetilde{j}_R & \widetilde{j}_L \end{array} \right\} \vcenter{\hbox{\scriptsize{
\setlength{\unitlength}{1cm}
\begin{picture}(4.4,2.4)
\put(0.6,0.9){\line(1,0){0.4}}
\put(0.6,1.5){\line(1,0){0.4}}
\put(0.6,0.9){\line(0,1){0.6}}
\put(1.0,0.9){\line(0,1){0.6}}
\put(0.7,1.1){$\Lambda$}
\put(0.8,1.5){\line(0,1){0.5}}
\put(0.8,0.4){\line(0,1){0.5}}
\put(0.8,2.0){\line(1,0){1.0}}
\put(0.8,0.4){\line(1,0){1.0}}
\put(2.0,2.0){\circle{0.4}}
\put(2.0,0.4){\circle{0.4}}
\put(1.9,1.9){T}
\put(1.9,0.3){T}
\put(2.0,1.4){\line(0,1){0.4}}
\put(2.0,0.6){\line(0,1){0.4}}
\put(2.1,1.5){$0 2$}
\put(2.1,0.7){$\frac{1}{2} 1$}
\put(2.2,0.4){\line(1,0){0.4}}
\put(2.2,2.0){\line(1,0){0.4}}
\put(2.7,1.9){$j_R N_R \alpha_R$}
\put(2.7,0.3){$\widetilde{j}_R N_R \widetilde{\alpha}_R$}
\put(0.0,2.1){$j_R (N_R-2) \alpha_L$}
\put(0.0,0.1){$\widetilde{j}_L (N_R - 1) \widetilde{\alpha}_L$}
\end{picture}
}}}
\end{align}
where the curly brackets denote Wigner 6-j symbols. The second example can hence also be decomposed in terms containing a structural part and a degeneracy part. The reduced tensors corresponding to the direct product of two spin $\frac{1}{2}$ operators are a spin $0$ tensor ($F^0$ in the example) and a spin $1$ tensor ($F^1$ in the example).
\end{widetext}

\bibliography{biblio}

\end{document}